\documentstyle[aps,preprint,prl,citesort,epsf]{revtex}

\tightenlines

\newcommand{\eqref}[1]{Eq.~(\protect\ref{#1})}

\begin{document}

\draft

\title{Unstable Periodic Orbit Analysis of Histograms of Chaotic Time Series}

\author{Scott M. Zoldi$^{\dagger}$}


\address{$^{\dagger}$Department of Physics and Center for Nonlinear and
Complex Systems, Duke University, Durham, North Carolina 27708}

\date{\today}
\maketitle


\begin{abstract}
Using the Lorenz equations, we have investigated whether
unstable periodic orbits (UPOs) associated with a strange
attractor may predict the occurrence of the robust sharp peaks
in histograms of some experimental chaotic time series. Histograms 
with sharp peaks occur for the Lorenz parameter value~$r=60.0$ but 
not for~$r=28.0$, and the sharp peaks for~$r=60.0$ do not correspond 
to any single histogram derived from a UPO. Despite this lack of 
correspondence, we show that histograms derived from a chaotic 
time series can be accurately predicted by an escape-time weighting 
of UPO histograms.
\end{abstract}


           

\narrowtext

Time series are often the only data available to quantify the complicated
aperiodic dynamics in experimental chaotic systems such as convecting
fluids, electronic circuits, and laser systems~\cite{reviews}.  Time series
derived from deterministic chaos exhibit structure under different
statistics which serves to characterize the chaotic state, e.g. the time
series of a laser experiment in a chaotic regime can have intensity
histograms with sharp reproducible peaks~\cite{Roy91}.  Chaotic time series
have also been used to detect unstable periodic orbits (UPOs) directly from
experimental data~\cite{detect}.  Given a knowledge of these UPOs, one can
characterize chaotic time series~\cite{characterize}, control
chaos~\cite{controlrefs}, approximate natural measures~\cite{invariant},
predict mean quantities~\cite{avepapers,Zoldi97KSUPOS}, and optimize
performance functions in the system~\cite{optimal}.  Unfortunately, it is a
difficult task to detect sufficiently many UPOs directly from time series
data to characterize chaos, causing the current reliance on the statistics
of time series to characterize chaotic states.

The aim of this paper is to examine the relative success and limitations of
trace formulas and escape-time weightings of UPOs in predicting the
structure of histograms of chaotic time series data.  To this end, the
Lorenz equations~\cite{Lorenz63} are well-suited as they can model a
single-mode laser~\cite{Haken75} and therefore be used to make qualitative
comparisons with chaotic multimode laser experiments that have 
intensity histograms that exhibit sharp reproducible peaks~\cite{Roy91}.
The structure of the x-variable histogram of the Lorenz equations
qualitatively changes from smooth for~$r=28.0$ to exhibiting sharp
reproducible peaks at~$r=60.0$ and hence these histograms characterize the
chaotic state.  Further in using UPOs to characterize chaos, the Lorenz
equations represents both ideal chaotic flows where the symbolic dynamics
is understood~($r=28.0$) and the more general non-ideal case where the
symbolic dynamics of the flow is not understood~($r=60.0$).  We demonstrate
that the structure of histograms of the chaotic x-variable can be predicted
in terms of the UPOs of the Lorenz equations using trace formulas where a
symbolic dynamics is understood and escape-time averages where a symbolic
dynamics is not understood.

The importance of unstable periodic orbits in understanding Axiom-A chaotic
systems has been well-known since the pioneering work of
Smale~\cite{Smale67}. Given a {\em complete} and {\em ordered} set of UPOs
up to some symbolic length~\cite{Dettman97}, trace formulas can be used to
approximate the natural measure of Axiom-A chaotic
attractors~\cite{invariant}.  Although impressive work has been done in
approximating averages using trace formulas and cycle expansions in
low-dimensional chaotic maps~\cite{avepapers,Artuso90}, much less is known
about chaotic flows in ordinary and partial differential equations that may
better describe many sorts of experimental chaotic systems.  Statistical
averages of UPOs based on trace formulas~\cite{invariant} and cycle
expansions~\cite{Artuso90} developed in the context of low-dimensional
chaotic maps become inapplicable to chaotic flows due to the lack of a
well-understood symbolic dynamics.  A unique symbolic dynamics is required
both to determine a complete set of low-period UPOs and to order these
UPOs.  As an example, in the high-fractal-dimension~$D=8.8$
Kuramoto-Sivashinsky equation (where the symbolic dynamics is not
understood) a simple escape-time weighting of UPOs proved superior to
approximate trace formulas in predicting averages~\cite{Zoldi97KSUPOS}.
Even for low-dimensional chaotic flows, one must identify a complete
symbolic dynamics of the flow and must compute a relatively complete set of
UPOs or trace formulas will not converge.  When these two conditions are
not satisfied, the escape-time weighting of UPOs may prove superior in
approximating averages of low-dimensional chaotic flows such as the
structure in histograms of chaotic time series.

The escape-time weighting (ETW) approximates averages of chaos without a
knowledge of symbolic dynamics and without complete ordered sets of
UPOs~\cite{Zoldi97KSUPOS}.  ETW weights the average quantity~$m_j$ computed
over UPO$_j$ by the escape-time $\tau_j$,
\begin{equation}
\tau_j  = {1 \over \sum_{+}\lambda_{ij}}
\end{equation} 
where the above sum is over all positive Lyapunov exponents~$\lambda_{ij}$
of UPO$_{j}$.  Since ETW is based on Lyapunov exponents, it is an {\bf
intensive} weighting which eliminates the necessity of ordering UPOs by
symbolic length or period.  Physically, the escape-time~$\tau_j$ of a UPO
is a time scale which reflects how long trajectories remain close to and
track a UPO before the trajectory diverges due to the instability of the
orbit.  Escape-times of UPOs are typically shorter than a UPO period
perhaps making ETW better suited for approximating averages in chaotic
systems that do not exhibit near recurrences over typical experimental
observation times.  Given that sufficiently many UPOs are computed, the
estimate of an average quantity~$m$ of chaos in terms of the same average
quantities~$m_j$ over UPOs is given by,
\begin{equation}
m = {{\sum \tau_j m_j} \over { \sum \tau_j}}.
\end{equation}
Table~\ref{logisticmap} demonstrates that the convergence of ETW can be
comparable to trace formulas~\cite{invariant} and zeta
functions~\cite{avepapers} even for low-dimensional maps with a known
symbolic dynamics such as the logistic map.  In addition, ETW is more
effective in predicting the structure of histograms of chaotic time series
as we discuss below.


To study the histogram structure of chaotic time series in terms of
computed UPOs we consider the Lorenz equations~\cite{Lorenz63},
\begin{equation}
\begin{array}{c}    
dx/dt= \sigma (y-x) \\ dy/dt = rx -y -xz \\ dz/dt=xy-bz
\end{array}                                                               
\end{equation}
where~$\sigma=10.0,~b=8/3$, and $r=28.0~{\rm or}~60.0$.  The Lorenz
equations were integrated with a 2nd-order accurate Adams-Bashforth method
with a time step of $dt=10^{-5}$ time units.  To construct long-time
histograms, the x-variable was recorded every time step over an integration
time of $10^8$ time units and the x-values were binned using 1000 equally
spaced bins that spanned the range of x from -20.0 to 20.0 for $r=28.0$ and
from -30.0 to 30.0 for $r=60.0$~(Fig.~\protect\ref{fig:hist-att}).
Histograms were not sensitive to modest changes in the number of bins.  The
histograms of long time series of the x-variable can be either smooth
($r=28.0$) or exhibit sharp reproducible peaks ($r=60.0$).  The number and
position of the sharp peaks for Lorenz parameters~$\sigma = 10.0,~b =
8/3,~{\rm and}~r=60.0$ are robust in the presence of large amounts of
multiplicative and additive noise suggesting that sharp reproducible peaks
in histograms may exist for many sorts of experimental time series.  For
$r=28.0$, short-time histograms (binning times $< 10^5$) exhibited random
placements of peaks dependent on initial condition, while for $r=60.0$
short-time histograms exhibited sharp peaks at the same characteristic
x-values similar to the experimental findings of Bracikowski et
al.~\cite{Roy91}.


To predict histograms such as those in~Fig.~\protect\ref{fig:hist-att} in
terms of UPOs, a damped-Newton algorithm~\cite{Zoldi97KSUPOS} was applied
to the Lorenz equations to compute 664 UPOs for~$r=28.0$ and 599 UPOs
for~$r=60.0$.  The damped-Newton algorithm had a 95\% success rate of
computing a UPO given an initial guess which consisted of a point on the
chaotic attractor and a randomly chosen period~$T<15.0$.  Each newly
computed UPO was compared with those already computed to form a unique set
of non-repeated UPOs.  The x-variable UPO histograms exhibit sharp
peaks~(Fig.~\protect\ref{fig:hist-UPOR28}) due to the turning points in the
orbits. Empirically we observe that no single UPO histogram can predict the
placement of peaks in the $r=60.0$ long-time chaotic x-variable histogram.
However, for binning times on the order of a UPO period, histograms of the
chaotic x-variable often had similar placements of peaks as the UPO
histograms~(Fig.~\protect\ref{fig:hist-UPOR28}), providing evidence that
the computed UPOs lie on the chaotic attractor.


	As no single UPO could predict the smooth long-time histogram of
the chaotic x-variable for~$r=28.0$, trace formula and escape-time averages
using many UPOs were used to approximate the long-time histogram.  In order
to apply a trace formula, the set of computed UPOs at $r=28.0$ were ordered
using a complete binary symbolic dynamics determined by a Lorenz
map~\cite{Sparrow82}.  Using symbolic dynamics, we determined that the
computed set of $r=28.0$ unstable periodic orbits was missing only seven
orbits up to symbolic length~$N=10$.  The $N=10$ trace formula applied to
x-variable UPO histograms (102 UPOs) resulted in an average histogram that
well approximated the long-time x-variable histogram of the Lorenz
attractor~(Fig~\protect\ref{fig:hist-UPOapprox28}(b)).  The escape-time
weighting average of x-variable UPO histograms (using no ordering) was also
satisfactory except at~$x=0$~(Fig~\protect\ref{fig:hist-UPOapprox28}(c)).
The trace formula is superior to escape-time weighting for Lorenz parameter
$r=28.0$, because of the formula understanding of the symbolic dynamics of
the flow.  It is remarkable that sharp peaked x-variable UPO histograms can
in both the trace formula and the escape-time weighting predict the smooth
long-time histogram at Lorenz parameter~$r=28.0$. The great predictive
power of averages of UPOs is demonstrated by this calculation in noting
that the~$N=10$ trace formula used only 102 UPOs of period $T<7.8$ time
units to predict the smooth long-time histogram of the chaotic attractor
which required $10^8$ time units to construct!


At Lorenz parameter $r=60.0$, the long-time x-variable histogram is
qualitatively different than at $r=28.0$ exhibiting sharp reproducible
peaks at specific values of the x-variable.  At $r=60.0$ the UPOs are only
approximately ordered by the binary symbolic dynamics of the Lorenz
map~\cite{Franceschini93,Sparrow82}.  Without a unique symbolic dynamics it
is impossible to determine whether we have computed a sufficiently complete
and ordered set of UPOs on which to apply a trace formula.  Using the
approximate symbolic dynamics and hence an approximate ordering of UPOs,
the trace formula using all UPO histograms of symbolic length $N=10$ (118
UPOs) predicts an average histogram that does not accurately approximate
the long-time histogram of the chaotic
x-variable~(Fig~\protect\ref{fig:hist-UPOapprox60}(b)).  Trace formulas of
other symbolic lengths also resulted in inaccurate approximations.  Prior
work using the same incomplete binary symbolic dynamics and trace
formalisms to estimate the Hausdorff dimension at Lorenz parameter $r=60.0$
demonstrated some success due to the insensitivity of this {\em particular
average quantity} to missing cycles~\cite{Franceschini93}. In general,
approximate trace formulas will not succeed in accurately predicting
averages as is demonstrated above by the~$r=60.0$ trace formula average
histogram.  Applying ETW to all {\em computed} x-variable UPO histograms
resulted in an average histogram that predicts the peaks in the long-time
histogram of the chaotic x-variable with incredible
accuracy~(Fig~\protect\ref{fig:hist-UPOapprox60}(c)) (the peak at $x=0$ is
due to a slight over-weighting homoclinic orbits).  ETW ignores the
ordering of UPOs and is also less sensitive to incomplete sets of UPOs (as
is the case for the~$r=60.0$ data) making ETW preferable in averaging UPOs
of chaotic flows.

Using trace formulas of unstable periodic orbits to predict averages in
Axiom-A chaotic systems is well-established for low-dimensional chaotic
flows which can be mapped onto a unique symbolic dynamics.  The agreement
between the trace formula average of UPO histograms at~$r=28.0$ and the
long-time histogram of the chaotic x-variable is an example of the accuracy
that can be achieved when a symbolic dynamics is understood.  For general
chaotic flows, trace formalisms will fail due to not understanding the
symbolic dynamics of the UPOs.  Not understanding the symbolic dynamics
causes both an inability to order the UPOs and an inability to determine
the completeness of a computed set of UPOs.  Escape-time weighting is an
intensive weighting {\bf not} based on symbolic dynamics and so does not
require any ordering of the UPOs making it preferable for approximating
averages of chaotic flows when the geometry of the attractor is not
understood.  As an example, the escape-time average of x-variable UPO
histograms at Lorenz parameter~$r=60.0$ predicts the reproducible peaks
found in the long-time histogram of the chaotic x-variable which can not be
predicted by trace formulas.  Not having a unique symbolic dynamics and not
having complete sets of UPOs will be general consequences of characterizing
low- and high-dimensional chaotic flows making ETW a powerful method for
extracting useful information from unstable periodic orbits of flows.

Helpful conversations with Predrag Cvitanovi\'c, Carl Dettmann, Henry
Greenside, and Ronnie Mainieri are acknowledged.  This work was
supported in part by the Computational Graduate Fellowship Program of
the Office of Scientific Computing in the Department of Energy, by
NSF grants NSF-DMS-93-07893 and NSF-CDA-92123483-04, and by DOE grant
DOE-DE-FG05-94ER25214.


\bibliographystyle{prsty}  



\newpage
\begin{figure}   
\caption{Long-time x-variable histogram~$\rho$ of the Lorenz
equations for {\bf (A)}: Lorenz parameters~$\sigma = 10.0,~b = 8/3,~{\rm
and}~r=28.0$ and for {\bf (B)}: Lorenz parameters~$\sigma = 10.0,~b =
8/3,~{\rm and}~r=60.0$.  Integration times of $10^8$ time units were 
used to construct the
histograms and 1000 bins were used to discretize the range of x between
-20.0 to 20.0 in {\bf (A)} and -30.0 to 30.0 in {\bf (B)}.}
\label{fig:hist-att}
\end{figure}

\begin{figure}   
\caption{Three representative x-variable histograms~$\rho$ of UPOs of the
Lorenz equations ($\sigma = 10.0,~b = 8/3,~{\rm
and}~r=28.0$): {\bf (A)}: histogram of UPO 011, {\bf (B)}: histogram of UPO
1000, {\bf (C)}: histogram of UPO 00010000000.  1000 bins were used to
divide the range of x between -20.0 to 20.0 }
\label{fig:hist-UPOR28}
\end{figure}

\begin{figure}   
\caption{{\bf (A)}: Long-time x-variable histogram~$\rho$ of the Lorenz
equations for Lorenz parameters~$\sigma = 10.0,~b = 8/3,~{\rm and}~r=28.0$.
{\bf (B)}: Symbolic length $N=10$ trace formula average histogram~$\rho_{\rm
TRACE}$ computed using 102 UPO histograms of the Lorenz equations. {\bf
(C)}: Escape-time weighting average histogram~$\rho_{\rm ESCAPE}$ using 664
{\em computed} UPOs of the Lorenz equations.  1000 bins were used to divide
the range of x between -20.0 to 20.0}
\label{fig:hist-UPOapprox28}
\end{figure}

\begin{figure}   
\caption{{\bf (A)}: Long-time x-variable histogram~$\rho$ of the Lorenz
equations for Lorenz parameters~$\sigma = 10.0,~b = 8/3,~{\rm and}~r=60.0$.
{\bf (B)}: Symbolic length $N=10$ trace formula average histogram~$\rho_{\rm
TRACE}$ computed using 118 UPO histograms of the Lorenz equations.  The
Lorenz map was used to construct a binary symbolic dynamics at $r=60.0$,
although not all UPOs had a unique symbolic designation.  {\bf (C)}:
Escape-time weighting average histogram~$\rho_{\rm ESCAPE}$ using 599 {\em
computed} UPOs of the Lorenz equations.  1000 bins were used to divide the
range of x between -30.0 to 30.0}
\label{fig:hist-UPOapprox60}
\end{figure}

\newpage
\centerline{\epsfysize=8.5in \epsfbox{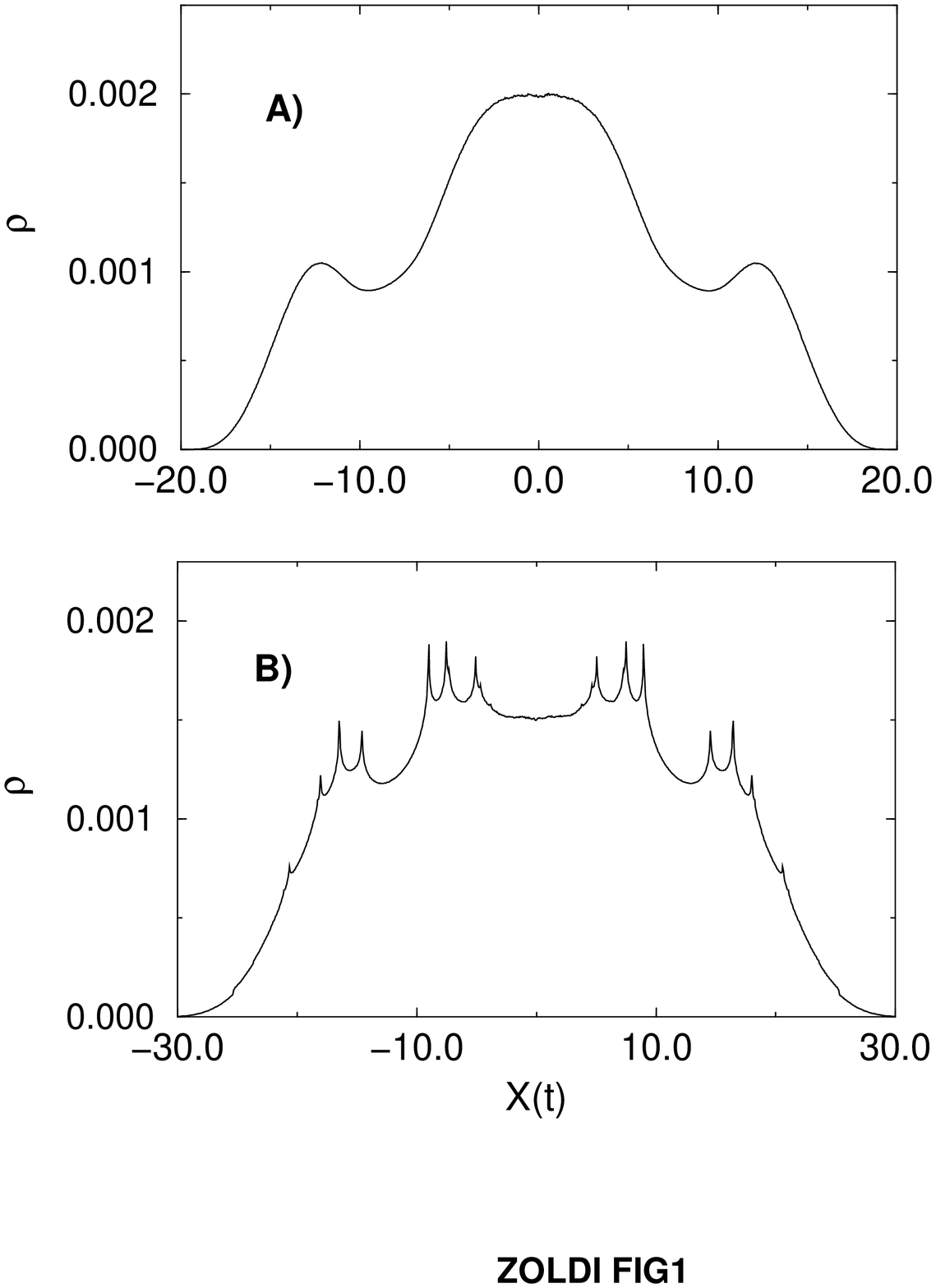}}

\newpage
\centerline{\epsfysize=8.5in \epsfbox{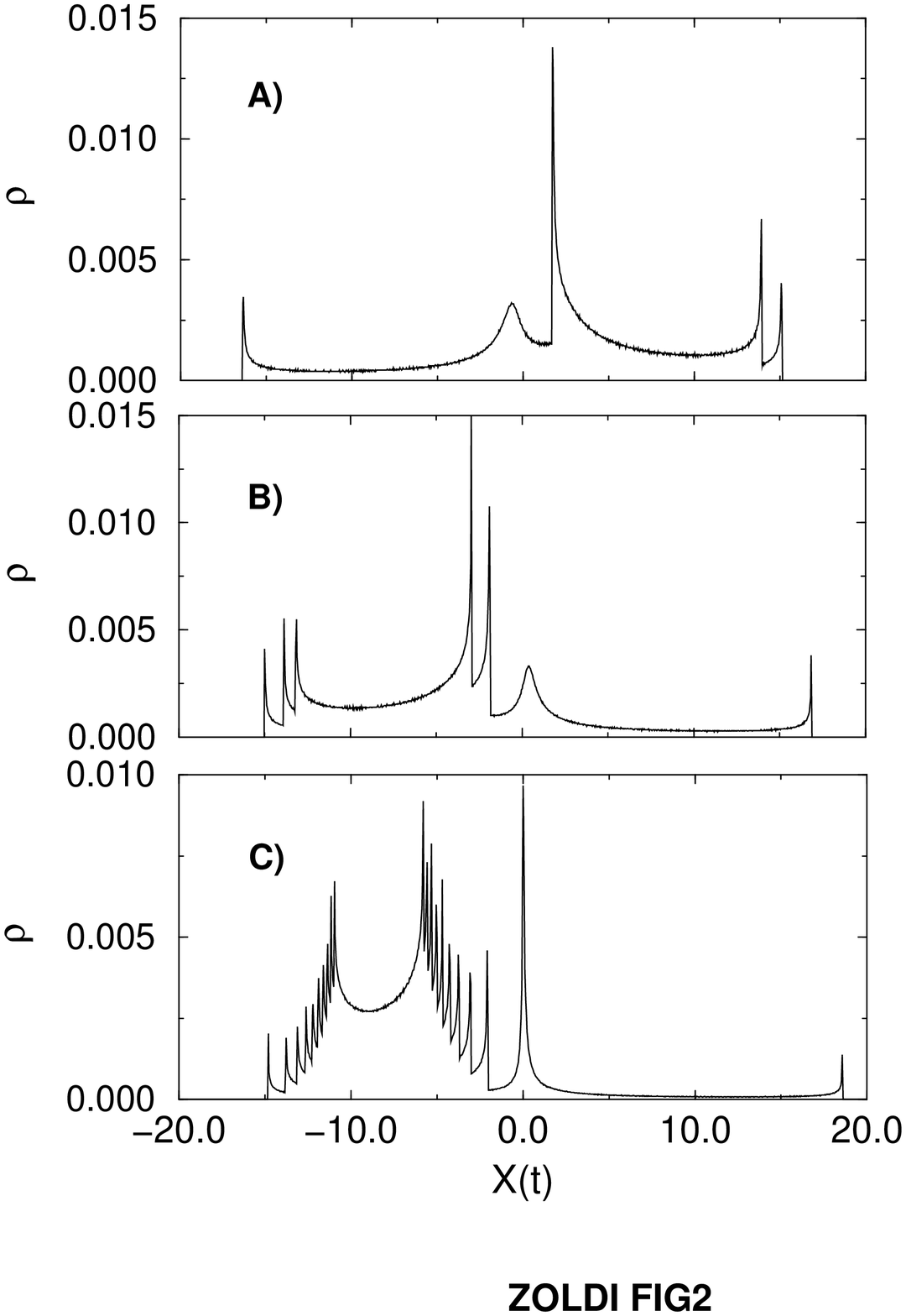}}

\newpage
\centerline{\epsfysize=8.5in \epsfbox{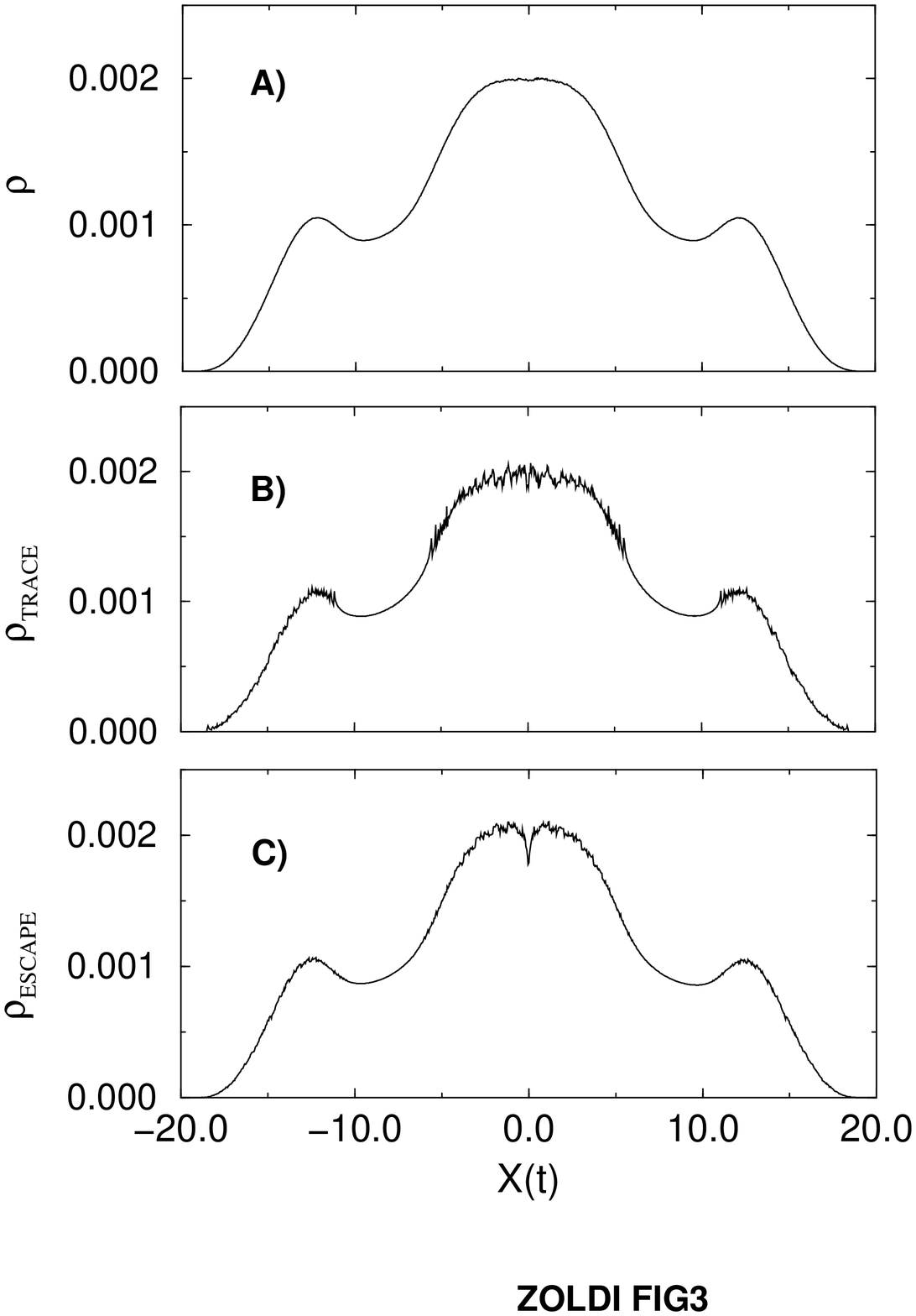}}

\newpage
\centerline{\epsfysize=8.5in \epsfbox{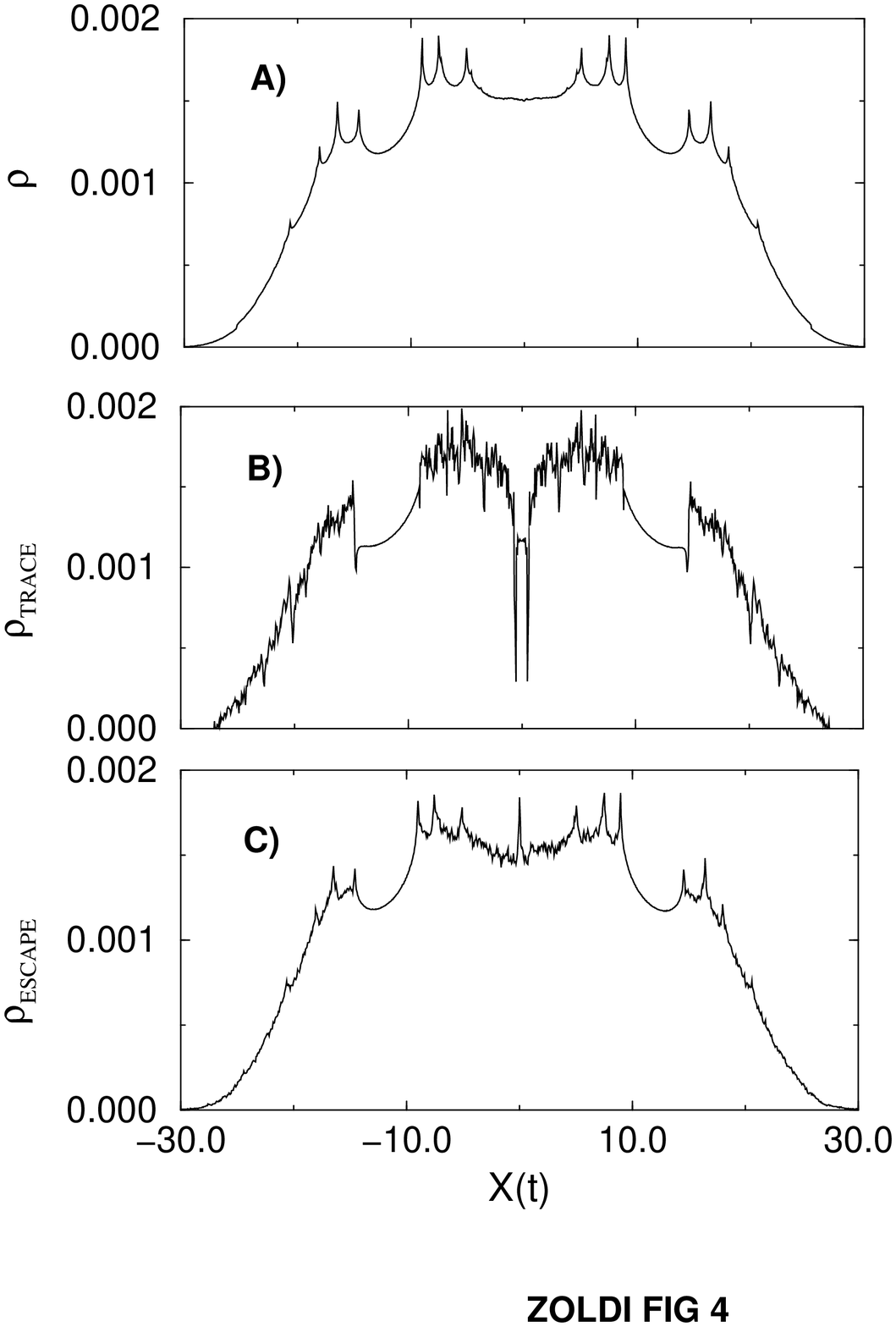}}

\twocolumn

\begin{table}
\caption{Approximations to the average 
of $<x^2>=0.375$ for the logistic map~$x[i+1]=4x[i](1-x[i])$ 
using complete sets of low-period UPOs.  $N$ denotes the
binary symbolic length of UPOs used in each 
approximation.}
\begin{tabular}{cdccdd}
   $N$ & ETW & TRACE & ZETA\\
\tableline
2 & 0.429 & 0.462 & 0.445 \\
3 & 0.414 & 0.421 & 0.424\\
4 & 0.392 & 0.398 & 0.406 
\end{tabular}
\label{logisticmap}
\end{table}

\end{document}